\documentclass[twocolumn,seceq]{jpsj3}
\usepackage{txfonts}
\usepackage{graphicx,bm,color,braket,here}

\title{Variational Approach to Many-Body Problems Incorporating Many-Body Effects at Finite Temperature}
\author{Akimitsu Kirikoshi, Wataru Kohno, and Takafumi Kita}
\inst{Department of Physics, Hokkaido University, Sapporo 060-0810, Japan}

\abst{We develop a variational approach at finite temperature that incorporates many-body correlation self-consistently.
The grand potential is constructed in terms of Green's function expressed by the variational parameters.
We apply this formalism to weakly interacting Bose-Einstein condensates to incorporate the dynamical 3/2-body processes, which are considered important in the dynamical properties.
The processes lower the free energy below the mean-field Hartree--Fock--Bogoliubov's value in the same way as a previous zero-temperature formalism.
From our numerical results, the pair creation or annihilation processes neglected in the Popov--Shohno approximation are enhanced, particularly in the long wavelength region, owing to the many-body effects.
Because the 3/2-body correlations give a finite contribution to the self-energy of quasiparticles, they may change the microscopic properties qualitatively, even in the weak-coupling region.} 

\begin{document}
\maketitle

\section{Introduction}

The use of a variational approach is one of the most powerful methods for studying many-particle systems.
In particular, some of the variational wave functions for the ground states have been constructed to consider correlations between particles or spins. \cite{Gutzwiller,Campbell,Laughlin,Brandow,Hang}
One of the most outstanding examples is the Gutzwiller wave function \cite{Gutzwiller}, which contains on-site correlations and has been used for the Hubbard model of electrons \cite{E1,E2} and bosons\cite{B1,B2,B3}.
At finite temperatures, the variational principle is given in terms of the grand potential as follows:\cite{Brown,mermin}
\begin{equation}\label{variational principle}
\Omega_{\rm{v}}=\braket{\hat{H}+\beta^{-1}\ln\hat{\rho}_{\rm{v}}}\geq\Omega;\ \beta=\frac{1}{k_BT}
\end{equation}
where $\Omega$ is the exact grand potential, and $\braket{\cdots}$ denotes a statistical average using the variational density matrix $\hat{\rho}_{\rm{v}}$.
There also exist some variational approaches at finite temperatures for incorporating correlations, particularly in strongly correlated electron systems \cite{Kakehashi} and quantum spin systems \cite{DMRG} because it is difficult to describe these systems using the standard mean-field approaches, e.g., Hartree-Fock theory.
On the other hand, most of the approaches adopt mean-field $\hat{\rho}_{\rm{v}}$.

The importance of the many-body effects is not limited to the strongly correlated system.
Although they may contribute to the thermodynamic and dynamic properties of the system, they are often neglected because it is difficult to incorporate them.
For example, in Bose-Einstein condensates (BECs), the dynamical 3/2-body processes, where one of two colliding non-condensed particles drops into the condensate and vice versa, are sources of many-body correlations.
The importance of these processes is emphasized in the dynamics of BECs at finite temperatures or near the critical temperature.\cite{Burnett,ZNG}
In recent years, however, we constructed a variational wave function for the ground state of weakly interacting bosons that self-consistently incorporates the dynamical 3/2-body processes.\cite{3/2,3/2m,3/2i}
Using it, we found that these processes contribute to the ground state energy with the same order as the mean-field energies.
Thus, the 3/2-body correlations should be incorporated self-consistently, even in the collisionless regime.

In this study, we develop a variational approach at finite temperatures to investigate the many-body effects in equilibrium states, which does not depend on the specific form of Hamiltonians.
We then apply it to the weak-coupling Bose-Einstein condensed phase and include the 3/2-body correlation self-consistently.
Unlike the zero-temperature formalism\cite{3/2,3/2m,3/2i}, it is difficult to evaluate the expectation values of Bogoliubov's quasiparticle operators.\cite{Bogoliubov,GA}
We evaluate them in terms of Green's function of these operators expressed by the variational parameters.

The remainder of this paper is organized as follows.
Section 2 constructs a variational density matrix with the 3/2-body correlation, and obtains an expression for the grand potential.
Section 3 presents numerical results for the free energy, condensate fraction, and pair correlation.
Section 4 summarizes this paper and describes some applications to other systems.
Appendix discusses the connection with the zero-temperature formalism.\cite{3/2}
Hereafter, we adopt the units $\hbar=k_B=1$.

\section{Formulation}

\subsection{Hamiltonian}

We consider a system of $\mathcal{N}$ identical bosons with mass $m$ and spin 0 in a box of volume $\mathcal{V}$ described by the grand-canonical Hamiltonian
\begin{equation}\label{Hamiltonian}
\hat{H}=\sum_{\bm k}(\varepsilon_k-\mu)\hat{c}_{\bm k}^\dag\hat{c}_{\bm k}+\frac{1}{2\mathcal{V}}\sum_{\bm k,\bm k',\bm q'}U_q\hat{c}_{\bm k+\bm q}^\dag\hat{c}_{\bm k'-\bm q}^\dag\hat{c}_{\bm k'}\hat{c}_{\bm k},
\end{equation}
where $\varepsilon_k\equiv k^2/2m$ is the kinetic energy, $\mu$ denotes the chemical potential, $(\hat{c}_{\bm k}^\dag,\hat{c}_{\bm k})$ are the field operators satisfying the Bose commutation relations, and $U_q$ is the interaction potential.
We aim to describe the equilibrium states of Eq. (\ref{Hamiltonian}) with Bose-Einstein condensation on the $\bm k=\bm0$ state.
Hence, it is convenient to classify $\hat{H}$ according to the number of non-condensed operators involved as
\begin{equation}\label{H}
\hat{H}=\hat{H}_0+\hat{H}_1+\hat{H}_{3/2}+\hat{H}_2.
\end{equation}
Each contribution on the right-hand side is given in terms of the primed sum $\displaystyle \sideset{}{'}\sum_{\bm k}\equiv\sum_{\bm k}\left(1-\delta_{\bm k,\bm 0}\right)$ as
\begin{subequations}\begin{align}
\hat{H}_0\equiv&-\mu\hat{c}_{\bm 0}^\dag\hat{c}_{\bm 0}+\frac{1}{2\mathcal{V}}U_0\hat{c}_{\bm 0}^\dag\hat{c}_{\bm 0}^\dag\hat{c}_{\bm 0}\hat{c}_{\bm 0},\\
\hat{H}_1\equiv&\sideset{}{'}\sum_{\bm k}(\varepsilon_k-\mu)\hat{c}_{\bm k}^\dag\hat{c}_{\bm k}+\frac{1}{\mathcal{V}}\sideset{}{'}\sum_{\bm k}(U_0+U_k)\hat{c}_{\bm 0}^\dag\hat{c}_{\bm 0}\hat{c}_{\bm k}^\dag\hat{c}_{\bm k}\notag\\
&+\frac{1}{2\mathcal{V}}\sideset{}{'}\sum_{\bm k}U_k\left(\hat{c}_{\bm k}^\dag\hat{c}_{-\bm k}^\dag\hat{c}_{\bm 0}\hat{c}_{\bm 0}+\hat{c}_{\bm 0}^\dag\hat{c}_{\bm 0}^\dag\hat{c}_{\bm k}\hat{c}_{-\bm k}\right),\\
\hat{H}_{3/2}\equiv&\frac{1}{\mathcal{V}}\sideset{}{'}\sum_{\bm k_1\bm k_2\bm k_3}\delta_{\bm k_1+\bm k_2+\bm k_3,\bm 0}U_{k_1}\left(\hat{c}_{\bm 0}^\dag\hat{c}_{-\bm k_3}^\dag\hat{c}_{\bm k_2}\hat{c}_{\bm k_1}+\hat{c}_{\bm k_1}^\dag\hat{c}_{\bm k_2}^\dag\hat{c}_{-\bm k_3}\hat{c}_{\bm 0}\right),\label{3/2body}\\
\hat{H}_2\equiv&\frac{1}{2\mathcal{V}}\sideset{}{'}\sum_{\bm k\bm k'\bm q}U_q\hat{c}_{\bm k+\bm q}^\dag\hat{c}_{\bm k'-\bm q}^\dag\hat{c}_{\bm k'}\hat{c}_{\bm k}.\label{2body}
\end{align}\end{subequations}

\subsection{Variational density matrix}
We select $\hat{\rho}_{\rm{v}}$ in terms of a variational Hamiltonian $\hat{H}_{\rm{v}}$ as
\begin{equation}\label{variational density matrix}
\hat{\rho}_{\rm{v}}=\exp\left[\beta\left(\Omega_{\rm{v,LW}}-\hat{H}_{\rm{v}}\right)\right]
\end{equation}
with
\begin{equation}\label{grand potential for Hv}
\Omega_{\rm{v,LW}}=-\frac{1}{\beta}\ln{\mathrm{Tr}e^{-\beta\hat{H}_{\rm{v}}}}.
\end{equation}
Substituting Eq. (\ref{variational density matrix}) into Eq. (\ref{variational principle}), we obtain an expression for the variational grand potential as
\begin{equation}
\Omega_{\rm{v}}=\mathrm{Tr}\hat{\rho}_{\rm{v}}(\hat{H}-\hat{H}_{\rm{v}})+\Omega_{\rm{v,LW}}.\label{O2}
\end{equation}
An appropriate choice of $\hat{H}_v$ is crucial for our variational study.
Here, we express $\hat{H}_{\rm{v}}$ as the sum of two distinct contributions:
\begin{equation}
\hat{H}_{\rm{v}}\equiv\hat{H}_{\rm{v},1}+\hat{H}_{\rm{v},3/2}.\label{effective}
\end{equation}
The terms on the right side are defined as
\begin{subequations}\begin{align}
\hat{H}_{\rm{v},1}\equiv&\sideset{}{'}\sum_{\bm k}E_{\bm k}\hat{\gamma}_{\bm k}^\dag\hat{\gamma}_{\bm k},\label{Hv1}\\
\hat{H}_{\rm{v},3/2}\equiv&\frac{1}{3!}\sideset{}{'}\sum_{\bm k_1\bm k_2\bm k_3}b_{\bm k_1\bm k_2\bm k_3}\left(\hat{\gamma}_{\bm k_1}^\dag\hat{\gamma}_{\bm k_2}^\dag\hat{\gamma}_{\bm k_3}^\dag+\hat{\gamma}_{\bm k_1}\hat{\gamma}_{\bm k_2}\hat{\gamma}_{\bm k_3}\right),\label{Hv3/2}
\end{align}\end{subequations}
where $\hat{\gamma}_{\bm k}$ denotes the quasiparticle operator
\begin{equation}
\hat{\gamma}_{\bm k}=u_{\bm k}\hat{c}_{\bm k}-v_{\bm k}\hat{c}_{-\bm k}^\dag,\label{quasi}
\end{equation}
with 
\begin{equation}
\left[\begin{array}{c}u_{\bm k}\\ v_{\bm k}\end{array}\right]\equiv\frac{1}{\sqrt{1-|\phi_{\bm k}|^2}}\left[\begin{array}{c}1\\ \phi_{\bm k}\end{array}\right],\ \ \ \phi_{\bm k}^*=\phi_{\bm k}=\phi_{-\bm k},
\end{equation}
so that $[\hat{\gamma}_{\bm k},\hat{\gamma}_{\bm k'}^\dag]=\delta_{\bm k,\bm k'}$ is satisfied.
The inverse of Eq. (\ref{quasi}) is given by
\begin{equation}
\hat{c}_{\bm k}=u_{\bm k}\hat{\gamma}_{\bm k}+v_{\bm k}\hat{\gamma}_{-\bm k}^\dag.\label{op}
\end{equation}
By definition, the coefficient $b_{\bm k_1\bm k_2\bm k_3}$ is real and symmetric with respect to every permutation among $(\bm k_1,\bm k_2,\bm k_3)$.
Using Eq. (\ref{op}), one can evaluate various expectations of the operators in Eq. (2.3) in terms of Eq. (\ref{variational density matrix}) with Eq. (2.8).
It is convenient for this purpose to introduce the following quantities for $\bm k\neq\bm0$:
\begin{subequations}\begin{align}
\rho_{\bm k}\equiv&\mathrm{Tr}\hat{\rho}_{\rm{v}}\hat{c}_{\bm k}^\dag\hat{c}_{\bm k}=u_{\bm k}^2\braket{\hat{\gamma}_{\bm k}^\dag\hat{\gamma}_{\bm k}}+|v_{\bm k}|^2\braket{\hat{\gamma}_{-\bm k}\hat{\gamma}_{-\bm k}^\dag}\notag\\
=&|v_{\bm k}|^2+u_{\bm k}^2\braket{\hat{\gamma}_{\bm k}^\dag\hat{\gamma}_{\bm k}}+|v_{\bm k}|^2\braket{\hat{\gamma}_{-\bm k}^\dag\hat{\gamma}_{-\bm k}},\\
F_{\bm k}\equiv&\mathrm{Tr}\hat{\rho}_{\rm{v}}\hat{c}_{\bm k}\hat{c}_{-\bm k}=u_{\bm k}v_{\bm k}(\braket{\hat{\gamma}_{\bm k}^\dag\hat{\gamma}_{\bm k}}+\braket{\hat{\gamma}_{-\bm k}\hat{\gamma}_{-\bm k}^\dag})\notag\\
=&u_{\bm k}v_{\bm k}(\braket{\hat{\gamma}_{\bm k}^\dag\hat{\gamma}_{\bm k}}+\braket{\hat{\gamma}_{-\bm k}^\dag\hat{\gamma}_{-\bm k}}+1),\\
W_{\bm k_1\bm k_2;\bm k_3}\equiv&\mathrm{Tr}\hat{\rho}_{\rm{v}}\hat{c}_{-\bm k_3}^\dag\hat{c}_{\bm k_2}\hat{c}_{\bm k_1}\notag\\
=&v_{-\bm k_3}^*u_{\bm k_2}u_{\bm k_1}\braket{\hat{\gamma}_{\bm k_3}^\dag\hat{\gamma}_{\bm k_2}^\dag\hat{\gamma}_{\bm k_1}^\dag}+u_{-\bm k_3}v_{\bm k_2}v_{\bm k_1}\braket{\hat{\gamma}_{-\bm k_3}\hat{\gamma}_{-\bm k_2}\hat{\gamma}_{-\bm k_1}}.
\end{align}\end{subequations}
We define the number of condensed particles as
\begin{equation}
\mathcal{N}_0\equiv\mathrm{Tr}\hat{\rho}_{\rm{v}}\hat{c}_{\bm 0}^\dag\hat{c}_{\bm 0}.
\end{equation}
Using this, we can express the expectations of the operator products in Eq. (2.3) as
\begin{subequations}\begin{align}
\mathrm{Tr}\hat{\rho}_{\rm{v}}\hat{c}_{\bm 0}^\dag\hat{c}_{\bm 0}^\dag\hat{c}_{\bm 0}\hat{c}_{\bm 0}\approx&\mathcal{N}_0^2,\\
\mathrm{Tr}\hat{\rho}_{\rm{v}}\hat{c}_{\bm 0}^\dag\hat{c}_{\bm 0}\hat{c}_{\bm k}^\dag\hat{c}_{\bm k}\approx&\mathcal{N}_0\rho_{\bm k},\\
\mathrm{Tr}\hat{\rho}_{\rm{v}}\hat{c}_{\bm 0}^\dag\hat{c}_{\bm 0}^\dag\hat{c}_{\bm k}\hat{c}_{-\bm k}\approx&\mathcal{N}_0F_{\bm k},\\
\mathrm{Tr}\hat{\rho}_{\rm{v}}\hat{c}_{\bm 0}^\dag\hat{c}_{-\bm k_3}^\dag\hat{c}_{\bm k_2}\hat{c}_{\bm k_1}\approx&\mathcal{\sqrt{N}}_0W_{\bm k_1\bm k_2;\bm k_3},\label{G3}\\
\mathrm{Tr}\hat{\rho}_{\rm{v}}\hat{c}_{\bm k_1'}^\dag\hat{c}_{\bm k_2'}^\dag\hat{c}_{\bm k_2}\hat{c}_{\bm k_1}\approx&(\delta_{\bm k_1',\bm k_1}\delta_{\bm k_2',\bm k_2}+\delta_{\bm k_1',\bm k_2}\delta_{\bm k_2',\bm k_1})\rho_{\bm k_1}\rho_{\bm k_2}\notag\\
&+\delta_{\bm k_1',-\bm k_2'}\delta_{\bm k_1,-\bm k_2}F_{\bm k_1'}F_{\bm k_1}.\label{G4}
\end{align}\end{subequations}
Because we consider the thermodynamic limit, we have approximated $(\hat{c}_{\bm 0}^\dag)^n\hat{c}_{\bm 0}^m$ as $\mathcal{N}_0^{(n+m)/2}$.
In deriving Eqs. (\ref{G3}) and (\ref{G4}), we have also used the fact that the numbers of differences between the $\hat{\gamma}_{\bm k}^\dag$ and $\hat{\gamma}_{\bm k}$ in any of their products should be multiples of 3 to give finite contributions in terms of Eq. (\ref{variational density matrix}) with Eq. (2.8).
Using Eq. (2.14), we obtain an expression for Eq. (\ref{O2}) as
\begin{equation}\begin{aligned}
\Omega_{\rm{v}}=&-\mu \mathcal{V}\bar{n}_0+\frac{\mathcal{V}}{2}U_0\bar{n}_0^2\\
&+\sideset{}{'}\sum_{\bm k}\left[\varepsilon_k-\mu+\bar{n}_0(U_0+U_k)\right]\rho_{\bm k}+\bar{n}_0\sideset{}{'}\sum_{\bm k}U_kF_{\bm k}\\
&+\frac{\sqrt{\mathcal{N}_0}}{\mathcal{V}}\sideset{}{'}\sum_{\bm k_1\bm k_2\bm k_3}\delta_{\bm k_1+\bm k_2+\bm k_3,\bm 0}U_{k_1}\left(W_{\bm k_1\bm k_2;\bm k_3}+W_{\bm k_1\bm k_2;\bm k_3}^*\right)\\
&+\frac{1}{2\mathcal{V}}\sideset{}{'}\sum_{\bm k\bm k'}\left[(U_0+U_{|\bm k-\bm k'|})\rho_{\bm k}\rho_{\bm k'}+U_{|\bm k-\bm k'|}F_{\bm k}F_{\bm k'}\right]\\
&-\sideset{}{'}\sum_{\bm k}E_{\bm k}\braket{\hat{\gamma}_{\bm k}^\dag\hat{\gamma}_{\bm k}}-\frac{1}{3!}\sideset{}{'}\sum_{\bm k_1\bm k_2\bm k_3}b_{\bm k_1\bm k_2\bm k_3}\biggl(\braket{\hat{\gamma}_{\bm k_1}^\dag\hat{\gamma}_{\bm k_2}^\dag\hat{\gamma}_{\bm k_3}^\dag}\\
&+\braket{\hat{\gamma}_{\bm k_1}\hat{\gamma}_{\bm k_2}\hat{\gamma}_{\bm k_3}}\biggr)+\Omega_{\rm{v,LW}},\label{O3}
\end{aligned}\end{equation}
with
\begin{equation}
\bar{n}_0\equiv\mathcal{N}_0/\mathcal{V}.
\end{equation}
It follows that the two basic expectations in Eq. (\ref{O3}) are derivable from Eq. (\ref{grand potential for Hv}) as
\begin{subequations}\begin{align}
\braket{\hat{\gamma}_{\bm k}^\dag\hat{\gamma}_{\bm k}}=&\frac{\delta\Omega_{\rm{v,LW}}}{\delta E_{\bm k}},\\
\braket{\hat{\gamma}_{\bm k_1}^\dag\hat{\gamma}_{\bm k_2}^\dag\hat{\gamma}_{\bm k_3}^\dag}=&\braket{\hat{\gamma}_{\bm k_1}\hat{\gamma}_{\bm k_2}\hat{\gamma}_{\bm k_3}}=\frac{1}{2}\frac{\delta\Omega_{\rm{v,LW}}}{\delta b_{\bm k_1\bm k_2\bm k_3}}.
\end{align}\end{subequations}

\subsection{Expression of $\Omega_{\rm{v,LW}}$ and basic expectations}
To obtain them, we introduce the quasiparticle Green's function $\mathcal{G}_{\bm k}(\tau)$ as
\begin{equation}
\mathcal{G}_{\bm k}(\tau_1-\tau_2)\equiv-\braket{\hat{T}_{\tau}\hat{\gamma}_{\bm k}(\tau_1)\hat{\gamma}_{\bm k}^\dag(\tau_2)},
\end{equation}
where $\hat{\gamma}_{\bm k}(\tau)\equiv e^{\tau\hat{H}_v}\hat{\gamma}_{\bm k}e^{-\tau\hat{H}_v},\ 
\hat{\gamma}_{\bm k}^\dag(\tau)\equiv e^{\tau\hat{H}_{\rm{v}}}\hat{\gamma}_{\bm k}^\dag e^{-\tau\hat{H}_{\rm{v}}}
$, and $\hat{T}_{\tau}$ is a time-ordered operator for imaginary time $\tau$.
It has the symmetry $\mathcal{G}_{\bm k}(\tau)=\mathcal{G}_{\bm k}(\tau+\beta)$, so we can expand it as
\begin{equation}
\mathcal{G}_{\bm k}(\tau)=\frac{1}{\beta}\sum_{n=-\infty}^{\infty}e^{-i\varepsilon_n\tau}\mathcal{G}_{\bm k}(i\varepsilon_n),\ \ \ \varepsilon_n\equiv2n\pi/\beta.\label{Matsubara}
\end{equation}

Equation (\ref{grand potential for Hv}) can be regarded as the grand potential for the system of quasiparticles described by the effective Hamiltonian $\hat{H}_{\rm{v}}$.
As shown by Luttinger and Ward,\cite{LW} it is expressible in terms of $\mathcal{G}_{\bm k}(i\varepsilon_n)$ as
\begin{equation}\begin{aligned}\label{LW functional}
\Omega_{\rm{v,LW}}[\mathcal{G}]=&\frac{1}{\beta}\sideset{}{'}\sum_{p}e^{i\varepsilon_n0_+}\left[\ln{\left[-\mathcal{G}_{\bm k}^{-1}(i\varepsilon_n)\right]}+\mathcal{S}_{\bm k}(i\varepsilon_n)\mathcal{G}_{\bm k}(i\varepsilon_n)\right]\\
&+\Phi_{\rm{v,LW}}[\mathcal{G}],
\end{aligned}\end{equation}
where $p=(\bm k,i\varepsilon_n)$.
$\Phi_{\rm{v,LW}}[\mathcal{G}]$ consists of all the skeleton diagrams in the simple perturbation expansion with respect to $\hat{H}_{{\rm{v}},3/2}$ for $\Omega_{\rm{v,LW}}$ with replacement of unperturbed Green's functions $\mathcal{G}^{(0)}$ with $\mathcal{G}$, and the self-energy $\mathcal{S}_{\bm k}(i\varepsilon_n)$ is derivable from it by
\begin{equation}
\mathcal{S}_{\bm k}(i\varepsilon_n)=-\beta\frac{\delta\Phi_{\rm{v,LW}}}{\delta\mathcal{G}_{\bm k}(i\varepsilon_n)}.\label{self}
\end{equation}
Factor $e^{i\varepsilon_n0_+}$ can be omitted for the self-energy corresponding to the interaction of Eq. (\ref{Hv3/2}).
Eq. (\ref{LW functional}) has an important property in that it is stationary:
\begin{equation}\label{condition}
\frac{\delta\Omega_{\rm{v,LW}}}{\delta\mathcal{G}_{\bm k}(i\varepsilon_n)}=0
\end{equation}
with respect to the variation in $\mathcal{G}_{\bm k}(i\varepsilon_n)$ that obeys Dyson's equation:
\begin{equation}\label{Dyson}
\mathcal{G}_{\bm k}(i\varepsilon_n)=\frac{1}{i\varepsilon_n-E_{\bm k}-\mathcal{S}_{\bm k}(i\varepsilon_n)}.
\end{equation}
The key quantity in the functional of Eq. (\ref{LW functional}) is $\Phi_{\rm{v,LW}}$. We adopt the approximation of the lowest order for it as
\begin{equation}\begin{aligned}
\Phi_{\rm{v,LW}}[\mathcal{G}]\approx&\frac{1}{3!\beta}\sideset{}{'}\sum_{\bm k_1\bm k_2\bm k_3}b_{\bm k_1\bm k_2\bm k_3}^2\int_0^\beta d\tau_1\int_0^\beta d\tau_2\\
&\times\mathcal{G}_{\bm k_1}(\tau_1-\tau_2)\mathcal{G}_{\bm k_2}(\tau_1-\tau_2)\mathcal{G}_{\bm k_3}(\tau_1-\tau_2).\label{Phitau}
\end{aligned}\end{equation}
Substituting in Eq. (\ref{Matsubara}), we can transform Eq. (\ref{Phitau}) into
\begin{equation}\begin{aligned}
\Phi_{\rm{v,LW}}[\mathcal{G}]=&\frac{1}{3!\beta^2}\sideset{}{'}\sum_{p_1p_2p_3}b_{\bm k_1\bm k_2\bm k_3}^2\delta_{n_1+n_2+n_3,0}\\
&\times\mathcal{G}_{\bm k_1}(i\varepsilon_{n_1})\mathcal{G}_{\bm k_2}(i\varepsilon_{n_2})\mathcal{G}_{\bm k_3}(i\varepsilon_{n_3}).\label{Phi}
\end{aligned}\end{equation}
The corresponding self-energy is calculated by Eq. (\ref{self}) to be
\begin{equation}
\mathcal{S}_{\bm k}(i\varepsilon_n)=-\frac{1}{2\beta}\sideset{}{'}\sum_{p_2p_3}b_{\bm k\bm k_2\bm k_3}^2\delta_{n+n_2+n_3,0}\mathcal{G}_{\bm k_2}(i\varepsilon_{n_2})\mathcal{G}_{\bm k_3}(i\varepsilon_{n_3}).\label{SE}
\end{equation}
From the stationary condition of Eq. (\ref{condition}), we can perform the differentiations of Eq. (2.17) with respect to the explicit dependences in Eq. (\ref{LW functional}) with Eq. (\ref{Phi}) as
\begin{subequations}\begin{align}
\braket{\hat{\gamma}_{\bm k}^\dag\hat{\gamma}_{\bm k}}=&-\frac{1}{\beta}\sum_ne^{i\varepsilon_n0_+}\mathcal{G}_{\bm k}(i\varepsilon_n),\label{nk}\\
\braket{\hat{\gamma}_{\bm k_1}\hat{\gamma}_{\bm k_2}\hat{\gamma}_{\bm k_3}}=&b_{\bm k_1\bm k_2\bm k_3}\frac{1}{\beta^2}\sum_{n_1n_2n_3}\delta_{n_1+n_2+n_3,0}\notag\\
&\times\mathcal{G}_{\bm k_1}(i\varepsilon_{n_1})\mathcal{G}_{\bm k_2}(i\varepsilon_{n_2})\mathcal{G}_{\bm k_3}(i\varepsilon_{n_3}).\label{n123}
\end{align}\end{subequations}

\subsection{Stationary conditions}

Now, we derive the stationary conditions of Eq. (\ref{O3}) with respect to $\mathcal{N}_0,\phi_{\bm k},E_{\bm k},$ and $b_{\bm k_1\bm k_2\bm k_3}$ along with the equation for the average particle number.
Below, we set $b_{\bm k_1\bm k_2\bm k_3}=b_{-\bm k_1-\bm k_2-\bm k_3}$.

First, the equation $\mathcal{N}=-\partial\Omega_{\rm{v}}/\partial\mu$ for the average particle number gives an expression for the particle density $\bar{n}\equiv\mathcal{N}/\mathcal{V}$ as
\begin{equation}\label{density}
\bar{n}=\bar{n}_0+\frac{1}{\mathcal{V}}\sideset{}{'}\sum_{\bm k}\rho_{\bm k}.
\end{equation}
Next, the stationary condition $\delta\Omega_{\rm{v}}/\delta\mathcal{N}_0=0$ yields an expression for the chemical potential as
\begin{equation}\begin{aligned}\label{chemical}
\mu=&U_0\bar{n}+\frac{1}{\mathcal{V}}\sideset{}{'}\sum_{\bm k}U_k(\rho_{\bm k}+F_{\bm k})\\
&+\frac{1}{\sqrt{\mathcal{N}_0}\mathcal{V}}\sideset{}{'}\sum_{\bm k_1\bm k_2\bm k_3}\delta_{\bm k_1+\bm k_2+\bm k_3,\bm 0}U_{k_1}W_{\bm k_1\bm k_2;\bm k_3}.
\end{aligned}\end{equation}

To calculate stationary conditions with respect to $(\phi_{\bm k},E_{\bm k},b_{\bm k_1\bm k_2\bm k_3})$, it is convenient to introduce the following quantities:
\begin{subequations}\begin{align}
\xi_{\bm k}\equiv&\frac{\delta\Omega_{\rm{v}}}{\delta\rho_{\bm k}}=\varepsilon_k-\mu+U_0\bar{n}+U_k\bar{n}_0+\frac{1}{\mathcal{V}}\sideset{}{'}\sum_{\bm k'}U_{|\bm k-\bm k'|}\rho_{\bm k'},\label{xi}\\
\Delta_{\bm k}\equiv&\frac{\delta\Omega_{\rm{v}}}{\delta F_{\bm k}}=\bar{n}_0U_k+\frac{1}{\mathcal{V}}\sideset{}{'}\sum_{\bm k'}U_{|\bm k-\bm k'|}F_{\bm k'},\label{Delta}\\
\lambda_{\bm k}\equiv&\frac{1-\phi_{\bm k}^2}{1+\braket{\hat{\gamma}_{\bm k}^\dag\hat{\gamma}_{\bm k}}+\braket{\hat{\gamma}_{-\bm k}^\dag\hat{\gamma}_{-\bm k}}}\frac{\delta\braket{\hat{H}_{3/2}}}{\delta\phi_{\bm k}}\notag\\
=&\frac{1}{1+\braket{\hat{\gamma}_{\bm k}^\dag\hat{\gamma}_{\bm k}}+\braket{\hat{\gamma}_{-\bm k}^\dag\hat{\gamma}_{-\bm k}}}\sideset{}{'}\sum_{\bm k_2\bm k_3}a_{\bm k_2\bm k_3;-\bm k}^{(0)}\braket{\hat{\gamma}_{\bm k}^\dag\hat{\gamma}_{\bm k_2}^\dag\hat{\gamma}_{\bm k_3}^\dag}\label{chi}
\end{align}\end{subequations}
and
\begin{subequations}\begin{align}
E_{\bm k}^{(0)}\equiv&\sum_{\sigma=\pm}\left(\frac{\delta\Omega_{\rm{v}}}{\delta\rho_{\sigma\bm k}}\frac{\delta\rho_{\sigma\bm k}}{\delta\braket{\hat{\gamma}_{\bm k}^\dag\hat{\gamma}_{\bm k}}}+\frac{\delta\Omega_{\rm{v}}}{\delta F_{\sigma\bm k}}\frac{\delta F_{\sigma\bm k}}{\delta\braket{\hat{\gamma}_{\bm k}^\dag\hat{\gamma}_{\bm k}}}\right)\notag\\
=&u_{\bm k}^2\xi_{\bm k}+v_{-\bm k}^2\xi_{-\bm k}+u_{\bm k}v_{\bm k}\Delta_{\bm k}+u_{-\bm k}v_{-\bm k}\Delta_{-\bm k},\label{Ek0}\\
a_{\bm k_1\bm k_2;-\bm k_3}^{(0)}\equiv&\frac{\sqrt{\mathcal{N}_0}}{\mathcal{V}}\delta_{\bm k_1+\bm k_2+\bm k_3,\bm 0}u_{\bm k_1}u_{\bm k_2}u_{\bm k_3}\left[(U_{k_1}+U_{k_2})(1+\phi_{\bm k_1}\phi_{\bm k_2}\phi_{\bm k_3})\right.\notag\\
&+(U_{k_3}+U_{k_1})(\phi_{\bm k_1}+\phi_{\bm k_3}\phi_{\bm k_2})\notag\\
&\left.+(U_{k_2}+U_{k_3})(\phi_{\bm k_2}+\phi_{\bm k_3}\phi_{\bm k_1})\right],\label{a1230}\\
b_{\bm k_1\bm k_2\bm k_3}^{(0)}\equiv&\frac{\sqrt{\mathcal{N}_0}}{\mathcal{V}}\delta_{\bm k_1+\bm k_2+\bm k_3,\bm 0}u_{\bm k_1}u_{\bm k_2}u_{\bm k_3}[(U_{k_1}+U_{k_2})(\phi_{\bm k_3}+\phi_{\bm k_1}\phi_{\bm k_2})\notag\\
&+(U_{k_3}+U_{k_1})(\phi_{\bm k_2}+\phi_{\bm k_3}\phi_{\bm k_1})\notag\\
&+(U_{k_2}+U_{k_3})(\phi_{\bm k_1}+\phi_{\bm k_2}\phi_{\bm k_3})].\label{b1230}
\end{align}\end{subequations}
With these preliminaries, the three conditions $\delta\Omega_{\rm{v}}/\delta\phi_{\bm k}=0,\ \delta\Omega_{\rm{v}}/\delta E_{\bm k}=0$, and $\delta\Omega_{\rm{v}}/\delta b_{\bm k_1\bm k_2\bm k_3}=0$ yield
\begin{subequations}\begin{align}
\phi_{\bm k}=&\frac{-\xi_{\bm k}+\sqrt{\xi_{\bm k}^2-\Delta_{\bm k}^2+\lambda_{\bm k}^2}}{\Delta_{\bm k}-\lambda_{\bm k}},\label{phi}\\
E_{\bm k}=&E_{\bm k}^{(0)},\label{Ek}\\
b_{\bm k_1\bm k_2\bm k_3}=&b_{\bm k_1\bm k_2\bm k_3}^{(0)}.\label{b123}
\end{align}\end{subequations}

We note that Eq. (\ref{condition}) is equivalent to $\delta\Omega_{\rm{v}}/\delta\mathcal{G}_{\bm k}(i\varepsilon_n)=0$ in our formalism.
This means that $\Omega_{\rm{v}}$ is also the functional of $\mathcal{G}_{\bm k}(i\varepsilon_n)$, and the dynamics of quasiparticles are optimized by the variational principle.
While $\mathcal{G}_{\bm k}$ and $\mathcal{S}_{\bm k}$ are represented by $\phi_{\bm k}, E_{\bm k}$, and $b_{\bm k_1\bm k_2\bm k_3}$, these parameters depend on Green's functions through the basic expectations.
Thus, we must determine $\phi_{\bm k}, E_{\bm k}, b_{\bm k_1\bm k_2\bm k_3}$, and $\mathcal{G}_{\bm k}$ self-consistently by solving Eqs. (\ref{phi})-(\ref{b123}) with Eqs. (\ref{Dyson}), (\ref{SE}), (\ref{nk}), (\ref{n123}), and (2.28)-(2.31).

\section{Numerical Results}

\subsection{Models and numerical procedures}
Numerical calculations were performed for the contact interaction potential $U_k=U$.\cite{LHY}
For convenience, we alternatively express $U$ as
\begin{equation}
U_k=U=\frac{4\pi a_U}{m}.
\end{equation}
A cutoff wavenumber $k_{\rm{c}}$ is introduced into every summation over $\bm k$ as
\begin{equation}
\sideset{}{'}\sum_{\bm k}\rightarrow\sideset{}{'}\sum_{\bm k}\theta(k_{\rm{c}}-k)
\end{equation}
to remove the ultraviolet divergence inherent in the potential.
The $s$-wave scattering length $a$ of the potential is obtained by
\begin{equation}
\frac{m}{4\pi a}=\frac{1}{U}+\int\frac{d^3k}{(2\pi)^3}\frac{\theta(k_{\rm{c}}-k)}{2\varepsilon_k},
\end{equation}
which yields
\begin{equation}
a=\frac{a_U}{1+2k_{\rm{c}}a_U/\pi}.
\end{equation}
We choose $k_{\rm{c}}$ such that $k_{\rm{c}}a_U\ll1$ is satisfied, i.e., $a\approx a_U$.

The units of energy and wavenumber of this system are defined by
\begin{equation}
\varepsilon_{U}\equiv U\bar{n},\ k_{U}\equiv\sqrt{2m\varepsilon_U},
\end{equation}
respectively.
They are used to transform Eq. (\ref{O3}) into the dimensionless form $\Omega_{\rm{v}}/\mathcal{N}\varepsilon_U$ for numerical calculations.
Sums over $\bm{k}$ are transformed into integrals as follows \cite{3/2}:
\begin{subequations}
\begin{align}
&\frac{1}{\mathcal{N}}\sideset{}{'}\sum_{\bm{k}}=8\left(\frac{2a_U^3\bar{n}}{\pi}\right)^{1/2}\int _{0}^{\tilde{k}_{\rm{c}}}d\tilde{k}\tilde{k}^{2},\\
&\frac{1}{\mathcal{N}}\sideset{}{'}\sum_{\bm{k}_{2},\bm{k}_{3}} \delta_{\bm{k}+\bm{k}_{2}+\bm{k}_{3},\bm{0}}\notag\\
=&8\left(\frac{2a_U^3\bar{n}}{\pi}\right)^{1/2}\frac{1}{2\tilde{k}}\int _{0}^{\tilde{k}_{\rm{c}}}d\tilde{k}_{2}\tilde{k}_{2}\int _{|\tilde{k}-\tilde{k}_{2}|}^{{\rm{min}}(\tilde{k}+\tilde{k}_{2},\tilde{k}_{\rm{c}})}d\tilde{k}_{3}\tilde{k}_{3},
\end{align}
\end{subequations}
where $\tilde{k}\equiv k/k_{U}$.
$a_U^3\bar{n}$ gives the density within its interaction range.
Here, we consider this coupling constant so that $a_U^3\bar{n}\ll1$ is satisfied, corresponding to dilute ultracold atomic gases.
The free energy $\mathcal{F}=\Omega+\mu\mathcal{N}$ is expressible as 
\begin{equation}\begin{aligned}
\mathcal{F}=&\frac{\mathcal{N}\varepsilon_U}{2}\Biggl[1+\left(\frac{128}{15\sqrt{\pi}}-\frac{4\sqrt{2}}{\sqrt{\pi}}\tilde{k}_{\rm{c}}-\frac{4\sqrt{2}}{\sqrt{\pi}\tilde{k}_{\rm{c}}}+2c_1\right)(a_U^3\bar{n})^{1/2}\\
&+2c_2a_U^3\bar{n}\Biggr]-TS.\label{free}
\end{aligned}\end{equation}
$c_1$ represents the contribution of $\braket{\hat{\gamma}_{\bm k}^\dag\hat{\gamma}_{\bm k}}$ from $\hat{H}_1$, while $c_2$ is the contribution of the 3/2-body and 2-body correlations.
The last term expresses an entropy defined by $S\equiv-\braket{\ln{\hat{\rho}_v}}=\beta(\braket{\hat{H}_{\rm{v}}}-\Omega_{\rm{v,LW}})$.

Numerical procedures are as follows.
We use the result of the Bogoliubov approximation\cite{Bogoliubov}
\begin{equation}
\phi_{\bm k}=-\frac{\varepsilon_k+\varepsilon_U-E_{\bm k}^{\rm{B}}}{\varepsilon_U},\ E_{\bm k}=E_{\bm k}^{\rm{B}}\equiv\sqrt{\varepsilon_k(\varepsilon_k+2\varepsilon_U)}
\end{equation}
and $b_{\bm k_1\bm k_2\bm k_3}=0$ as initial solutions for the self-consistent equations at finite temperatures.

We now make some comments on the summations over Matsubara frequency $\varepsilon_n$.
First, some of them can be easily done computed using an imaginary time representation.
For example,
\begin{equation*}\begin{aligned}
\mathcal{S}_{\bm k}(\tau)\equiv&\frac{1}{\beta}\sum_{n}\mathcal{S}_{\bm k}(i\varepsilon_n)e^{-i\varepsilon_n\tau}\notag\\
=&-\frac{1}{2}\sideset{}{'}\sum_{\bm k_2\bm k_3}b_{\bm k\bm k_2\bm k_3}^2\mathcal{G}_{\bm k_2}(\beta-\tau)\mathcal{G}_{\bm k_3}(\beta-\tau).
\end{aligned}\end{equation*}
We use the fast Fourier transform (FFT)\cite{numerical} to evaluate the Fourier transform of $\mathcal{G}(i\varepsilon_n)$ into $\mathcal{G}(\tau)$.
This can also be applied to the transform $\mathcal{S}(\tau)\to\mathcal{S}(i\varepsilon_n)$ by discretizing $\tau$.
To perform these calculations, we must choose a cutoff frequency $\varepsilon_{n_{\rm{c}}}$.
We should note that the FFT of $\mathcal{G}_{\bm k}(i\varepsilon_n)$ for $k\gg k_U$ may suffer from substantial numerical errors.
However, it can be expected that the quasiparticles behave as the ideal bosons $E_{\bm k}\approx\varepsilon_k$ in the region $k\gg k_U$, so we can approximate $\mathcal{G}_{\bm k}(i\varepsilon_n)\approx\mathcal{G}_{\bm k}^{(0)}(i\varepsilon_n)\approx(i\varepsilon_n-\varepsilon_k)^{-1}$ for $k\lesssim k_{\rm{c}}$.
Because $|\mathrm{Re}\mathcal{G}^{(0)}_{\bm k}(i\varepsilon_n)|\approx\varepsilon_k(\varepsilon_n^{2}+\varepsilon_k^2)^{-1}$ and $|\mathrm{Im}\mathcal{G}_{\bm k}^{(0)}(i\varepsilon_n)|\approx\varepsilon_n(\varepsilon_n^{2}+\varepsilon_k^2)^{-1}$, it is sufficient to choose $\varepsilon_{n_{\rm{c}}}$ to satisfy $\varepsilon_{n_{\rm{c}}}\gg\varepsilon_{k_{\rm{c}}}$.
We fixed the number of integration points for Matsubara frequency and imaginary time at $n_{\rm{c}}=2^{14}\sim10^4$.
Secondly, we must obtain the value of the Matsubara Green's function at the discontinuity point $\tau=0_{\pm}$ from the FFT.
It is known that the Fourier series $\tilde{f}_{\rm{FT}}(x)$ converges to $\frac{1}{2}\left[f(a+0_+)+f(a+0_-)\right]$ at the discontinuity point $x=a$.
Because, $\mathcal{G}_{\bm k}(\tau=0_+)-\mathcal{G}_{\bm k}(\tau=0_-)=-\braket{\hat{\gamma}_{\bm k}\hat{\gamma}_{\bm k}^\dag}+\braket{\hat{\gamma}_{\bm k}^\dag\hat{\gamma}_{\bm k}}=-1$, $\braket{\hat{\gamma}_{\bm k}^\dag\hat{\gamma}_{\bm k}}$ is expressed by
\begin{equation}
\braket{\hat{\gamma}_{\bm k}^\dag\hat{\gamma}_{\bm k}}=-\mathcal{G}(\tau=0_-)=-\tilde{\mathcal{G}}_{\rm{FT}}(0)-\frac{1}{2}.
\end{equation}
Finally, because it is difficult to determine the imaginary time dependence, we evaluate the first term of Eq.(\ref{LW functional}) numerically by substituting $0_+=10^{-30}/\varepsilon_{n_{\rm{c}}}$.
The convergence of the iteration can be checked by monitoring the grand potential energy $\Omega_{\rm{v}}$.
We stopped the iteration when the magnitude of the relative difference between the old and new grand potentials decreased to below $10^{-10}$.

\subsection{Results}
In this section, we compare the results obtained by the effective Hamiltonian $\hat{H}_{\rm{v,1}}+\hat{H}_{\rm{v,3/2}}$ with those of $\hat{H}_{\rm{v}}=\hat{H}_{\rm{v,1}}$, i.e., the Hartree--Fock--Bogoliubov (HFB) approximation.\cite{HFB}
Because at the low temperatures, the 3/2-body correlations are considered to be the main sources of the many-body effects compared to the collisions between non-condensed particles, i.e., the dynamical 2-body correlations, we solved the self-consistent equations in the region $0< T<T_{\rm{c0}}/2$, where $T_{\rm{c0}}$ is the critical temperature for an ideal system.

\begin{figure}[t]
\begin{center}
\includegraphics[width=1.0\linewidth]{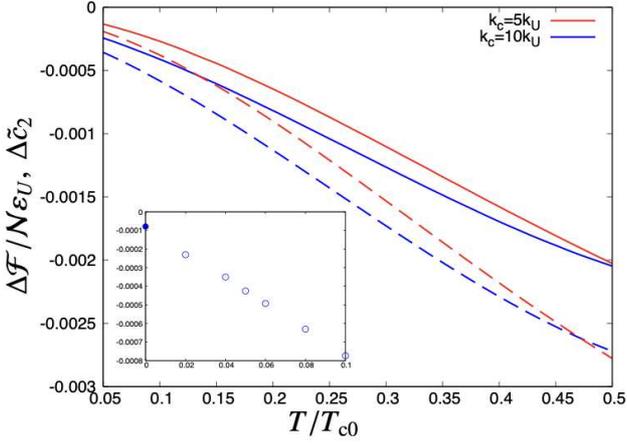}
\end{center}
\caption{(Color online)
Temperature dependencies of $\Delta\mathcal{F}$ (solid) and $\Delta\tilde{c}_2$ (dashed), where $\Delta\tilde{c}_2\equiv\Delta c_2\times a_U^3\bar{n}$ for $a_U^3\bar{n}=10^{-6}$.
The inset shows $\Delta\mathcal{F}$ near zero temperature for $k_{\rm{c}}=10k_U$, and the filled circle is obtained by zero-temperature formalism.\cite{3/2}}
\label{free1}
\end{figure}

\begin{figure}[t]
\begin{center}
\includegraphics[width=1.0\linewidth]{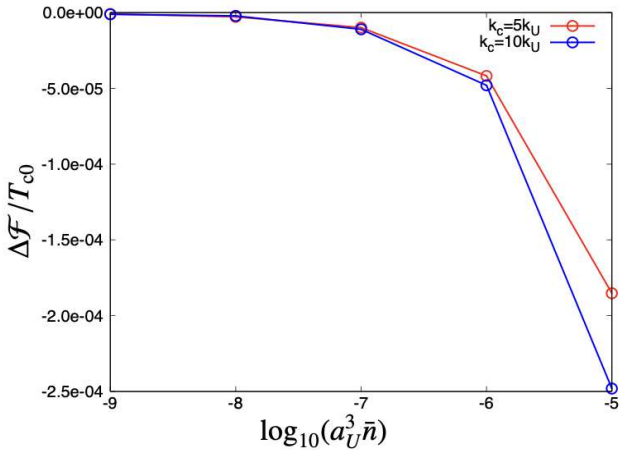}
\end{center}
\caption{(Color online)
$\Delta\mathcal{F}$ as functions of $\log_{10}(a_U^3\bar{n})$ for $T=0.3T_{\rm{c0}}$.
}
\label{free2}
\end{figure}

First, we estimate the free energy per particle.
We set the energy difference from the HFB theory as $\Delta\mathcal{F}\equiv\mathcal{F}-\mathcal{F}^{\rm{HFB}}$, where $\mathcal{F}^{\rm{HFB}}$ represents the free energy of the HFB approximation.
Figure \ref{free1} shows the $T$ dependence of the deviations of the free energy $\Delta\mathcal{F}$ and the coefficient $\Delta c_2\equiv c_2-c_2^{\rm{HFB}}$.
We observe that  the free-energy is lower than $\mathcal{F}^{\rm{HFB}}$ ($\Delta\mathcal{F}<0$), and it decreases by an order of $a_U^3\bar{n}$.
However, the deviations of $c_1$ and the entropy depend on the cutoff in our calculation.
The inset in Fig. \ref{free1} shows that our calculation is numerically connected to the result of zero temperature formalism\cite{3/2}.
Figure \ref{free2} plots $\Delta\mathcal{F}$ as functions of $\log_{10}(a_U^3\bar{n})$.
These results show that the free energy is reduced by enhancement of the many-body effect with density and temperature.
In summary, the mean-field theory, which is the standard extension of Bogoliubov theory\cite{Bogoliubov} to finite temperatures, is not effective for describing the weak-coupling BEC as well as the zero-temperature system, so the 3/2-body correlations should be incorporated self-consistently to investigate the thermodynamic properties, even in a collisionless regime.

\begin{figure}[t]
\begin{center}
\includegraphics[width=1.0\linewidth]{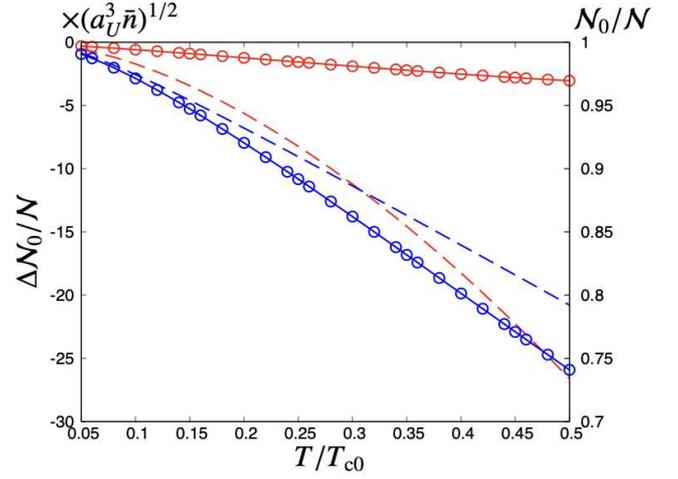}
\end{center}
\caption{(Color online)
Shifts of condensate fraction from the results of the HFB theory for $a_U^3\bar{n}=1.0\times10^{-6}$ (red) or $1.0\times10^{-9}$ (blue) and $k_{\rm{c}}=10k_U$.
Dashed lines represent the number of condensed particles $\mathcal{N}_0$.
}
\label{condensate}
\end{figure}

\begin{figure}[t]
\begin{center}
\includegraphics[width=1.0\linewidth]{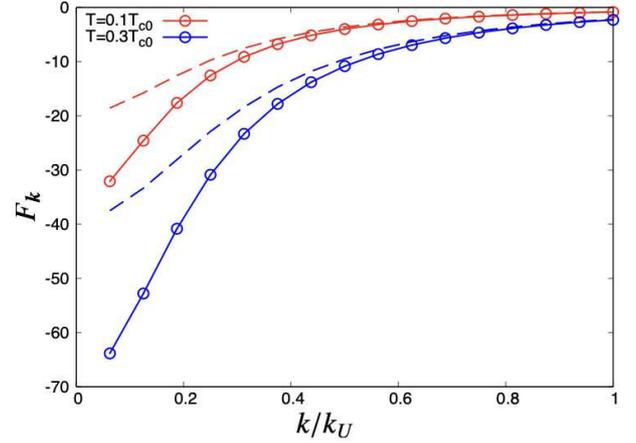}
\end{center}
\caption{(Color online)
Plot of $F_{\bm k}$ as functions of $k$ compared with the results of the HFB theory (dashed lines) for $k_{\rm{c}}=10k_U$ and $a_U^3\bar{n}=10^{-6}$.
}
\label{pair}
\end{figure}

To see the contributions of the many-body effects to the thermodynamic properties, we calculate the condensate fractions from Eq. (\ref{density}).
Figure \ref{condensate} shows that the number of condensed particle decreases due to the 3/2-body correlations: $\Delta\mathcal{N}_0\equiv\mathcal{N}_0-\mathcal{N}_0^{\rm{HFB}}<0$.
Because the 3/2-body correlations give the opportunity for particle exchange between condensate and non-condensate, they tend to enhance other correlations.
Indeed, the pair correlations $F_{\bm k}=\mathrm{Tr}\hat{\rho}_{\rm{v}}\hat{c}_{\bm k}\hat{c}_{-\bm k}$ are enhanced due to the 3/2-body correlations, particularly in the long wavelength region, as shown in Fig. \ref{pair}.
This indicates that $F_{\bm k}$ should be incorporated self-consistently in BEC phase.
Therefore, the Popov--Shohno approximation\cite{Popov1,Shohno}, which is obtained by $\hat{H}_{\rm{v}}=\hat{H}_{\rm{v,1}}$ with $F_{\bm k}$ neglected to give a gapless excitation, is invalid even in the collisionless regime.
This can also be confirmed from the fact that the free-energy of the Popov--Shohno approximation is higher than that of the HFB approximation; therefore, the pair creation or annihilation processes also contribute to the stabilization of the BEC system.

\section{Summary}
We have constructed a variational density matrix incorporating the many-body effects self-consistently.
In general, if the trial density matrix includes the interaction terms, it is difficult to calculate the right hand side of Eq. (\ref{grand potential for Hv}) directly.
The advantage of our formalism is that by considering $\Omega_{\rm{v,LW}}$ as the Luttinger--Ward functional and using Green's function, it is possible to evaluate the basic expectations and derive Dyson's equation that determines quasiparticle dynamics from the variational principle.
We adopted this formalism in a BEC system, which incorporates the dynamical 3/2-body processes to give a lower free energy than the mean-field density matrix.
Therefore, it turns out that the particle correlations contribute to stabilizing the system.
Because the 3/2-body correlations provide the collisions between the condensate and non-condensate, they also contribute to increasing entropy.
However, we could not confirm this through our numerical results.

A qualitative difference from the mean-field theory cannot be found in the above thermodynamic properties.
On the other hand, the self-energy of the quasiparticles $\mathcal{S}_{\bm k}(i\varepsilon_n)$ becomes finite due to the 3/2-body correlations.
Since this quantity contributes to the spectrum of excitations in the form of an energy shift or width of the spectrum, qualitative changing of the single-particle excitation could occur, as discussed in the zero-temperature formalism.\cite{3/2}
Thus, the many-body effects may  bring about qualitative changing of the microscopic properties from the mean-field theory, even in a collisionless regime.
We intend to investigate this practically in the near future.

In this paper, we only considered the case of a single-component BEC phase at low temperatures.
Our formalism can be used at higher temperatures, where the dynamical 2-body correlations are more important.
In this case, only $\hat{H}_{\rm{v}}$ must be constructed so as to characterize these processes.
It can also be adopted in other systems such as the superconductor and Bose-Fermi mixture systems as follows:
\begin{enumerate}
\item[(i)] We construct $\hat{\rho}_{\rm{v}}$ that includes the terms considered to be the sources of the many-body effects in the systems.
\item[(ii)] $\Omega_{\rm{v,LW}}$ is evaluated as a functional of Green's function, and $\Phi_{\rm{v,LW}}$ determines the many-body effects.
\item[(iii)] Because $\Omega_{\rm{v}}$ is expressible using the variational parameters, we determine them from Eq. (\ref{variational principle}).
\end{enumerate}
It seems to be simple to extend to an inhomogeneous system such as the trapped system by using a general quantum number $q$ instead of the wavenumber $\bm k$ and Green's function in coordinate space.
Therefore, our variational approach can be used to study the many-body effects on the thermodynamic properties in other systems.

\acknowledgment
\begin{acknowledgment}
W. K. is a JSPS Research Fellow, and this work was supported in part by KAKENHI Grand Number 18J13241.
The numerical calculations were performed using the facilities of the Supercomputer Center, Institute for Solid State Physics, University of Tokyo.
We would like to thank Editage (www.editage.jp) for English language editing.
\end{acknowledgment}

\appendix
\section{Connection to the Zero-Temperature Formalism}
We adopt the Lehman representation for Green's function:
\begin{equation}
\mathcal{G}_{\bm k}(i\varepsilon_n)=\int_{-\infty}^\infty\frac{d\varepsilon}{2\pi}\frac{\mathcal{A}_{\bm k}(\varepsilon)}{i\varepsilon_n-\varepsilon},\label{GL}
\end{equation}
where $\mathcal{A}_{\bm k}(\varepsilon)$ satisfies the sum rule
\begin{equation}
\int_{-\infty}^{\infty}\frac{d\varepsilon}{2\pi}\mathcal{A}_{\bm k}(\varepsilon)=1.
\end{equation}
Using Eq. (\ref{GL}), we obtain an alternative expression for $\Phi_{\rm{v,LW}}[\mathcal{G}]$ and $\mathcal{S}_{\bm k}(i\varepsilon_n)$ as
\begin{equation}\begin{aligned}
\Phi_{\rm{v,LW}}=&-\frac{1}{3!}\sideset{}{'}\sum_{\bm k_1\bm k_2\bm k_3}|b_{\bm k_1\bm k_2\bm k_3}|^2\int_{-\infty}^{\infty}\frac{d\varepsilon_1}{2\pi}\int_{-\infty}^{\infty}\frac{d\varepsilon_2}{2\pi}\int_{-\infty}^{\infty}\frac{d\varepsilon_3}{2\pi}\\
&\times\frac{\mathcal{A}_{\bm k_1}(\varepsilon_1)\mathcal{A}_{\bm k_2}(\varepsilon_2)\mathcal{A}_{\bm k_3}(\varepsilon_3)}{\varepsilon_1+\varepsilon_2+\varepsilon_3}\\
&\times\left[\prod_{i=1}^3[1+f(\varepsilon_i)]-\prod_{i=1}^3f(\varepsilon_i)\right],
\end{aligned}\end{equation}
\begin{equation}\begin{aligned}
\mathcal{S}_{\bm k}(i\varepsilon_n)=&-\frac{1}{2}\sideset{}{'}\sum_{\bm k_2\bm k_3}b_{\bm k\bm k_2\bm k_3}^2\int_{-\infty}^{\infty}\frac{d\varepsilon_2}{2\pi}\int_{-\infty}^{\infty}\frac{d\varepsilon_3}{2\pi}\\
&\times\frac{\mathcal{A}_{\bm k_2}(\varepsilon_2)\mathcal{A}_{\bm k_3}(\varepsilon_3)}{i\varepsilon_n+\varepsilon_2+\varepsilon_3}[1+f(\varepsilon_2)+f(\varepsilon_3)].\label{SE2}
\end{aligned}\end{equation}
Here, $f(\varepsilon)$ denotes the Bose distribution function.
It is convenient to express Eq. (\ref{SE2}) in the Lehman representation too as
\begin{equation}
\mathcal{S}_{\bm k}(i\varepsilon_n)=\int_{-\infty}^{\infty}\frac{d\varepsilon}{2\pi}\frac{\varGamma_{\bm k}(\varepsilon)}{i\varepsilon_n-\varepsilon}.\label{LSE}
\end{equation}
$\varGamma_{\bm k}(\varepsilon)\equiv-2\rm{Im}\mathcal{S}_{\bm k}(\varepsilon+i0_+)$ is given explicitly by
\begin{equation}\begin{aligned}
\varGamma_{\bm k}(\varepsilon)=&-\frac{1}{2}\sideset{}{'}\sum_{\bm k_2\bm k_3}b_{\bm k\bm k_2\bm k_3}^2\int_{-\infty}^{\infty}\frac{d\varepsilon_2}{2\pi}\mathcal{A}_{\bm k_2}(\varepsilon_2)\mathcal{A}_{\bm k_3}(-\varepsilon-\varepsilon_2)\\
&\times[1+f(\varepsilon_2)+f(-\varepsilon-\varepsilon_2)].\label{LSE2}
\end{aligned}\end{equation}
The retarded self-energy is obtained from Eq. (\ref{LSE}) by $\mathcal{S}_{\bm k}^{\rm{R}}(\varepsilon)=\mathcal{S}_{\bm k}(\varepsilon+i0_+)$ as
\begin{equation}
\mathcal{S}_{\bm k}^{\rm{R}}(\varepsilon)=\frac{\rm{P}}{2\pi}\int_{-\infty}^{\infty}\frac{\varGamma_{\bm k}(\epsilon)}{\varepsilon-\epsilon}d\epsilon-\frac{i}{2}\varGamma_{\bm k}(\varepsilon),\label{RSE}
\end{equation}
where $\rm{P}$ denotes the principal value.
Using Eqs. (\ref{Dyson}) and (\ref{GL}), we obtain the spectral function $\mathcal{A}_{\bm k}(\varepsilon)=-2\rm{Im}\mathcal{G}_{\bm k}(\varepsilon+i0_+)$ as
\begin{equation}
\mathcal{A}_{\bm k}(\varepsilon)=\frac{\varGamma_{\bm k}(\varepsilon)}{[\varepsilon-E_{\bm k}-{\rm{Re}}\mathcal{S}_{\bm k}^{\rm{R}}(\varepsilon)]^2+[\varGamma_{\bm k}(\varepsilon)/2]^2}.\label{A}
\end{equation}
The quantities in Eqs. (\ref{nk}) and (\ref{n123}) are expressible as
\begin{subequations}\begin{align}
\braket{\hat{\gamma}_{\bm k}^\dag\hat{\gamma}_{\bm k}}=&\frac{1}{2\pi}\int_{-\infty}^{\infty}d\varepsilon\mathcal{A}_{\bm k}(\varepsilon)f(\varepsilon),\label{nkL}\\
\braket{\hat{\gamma}_{\bm k_1}\hat{\gamma}_{\bm k_2}\hat{\gamma}_{\bm k_3}}=&-b_{\bm k_1\bm k_2\bm k_3}\int_{-\infty}^{\infty}\frac{d\varepsilon_1}{2\pi}\int_{-\infty}^{\infty}\frac{d\varepsilon_2}{2\pi}\int_{-\infty}^{\infty}\frac{d\varepsilon_3}{2\pi}\notag\\
&\times\frac{\mathcal{A}_{\bm k_1}(\varepsilon_1)\mathcal{A}_{\bm k_2}(\varepsilon_2)\mathcal{A}_{\bm k_3}(\varepsilon_3)}{\varepsilon_1+\varepsilon_2+\varepsilon_3}\notag\\
&\times\left[\prod_{i=1}^3[1+f(\varepsilon_i)]-\prod_{i=1}^3f(\varepsilon_i)\right].\label{n123L}
\end{align}\end{subequations}

Here, we show that our formalism connects with the zero-temperature formalism.\cite{3/2}
First, we substitute the zeroth-order of the spectral function
\begin{equation}
\mathcal{A}_{\bm k}^{(0)}(\varepsilon)=-2{\rm{Im}}\mathcal{G}_{\bm k}^{(0)}(\varepsilon+i0_+)=2\pi\delta(\varepsilon-E_{\bm k}),\label{A0}
\end{equation}
into Eq. (\ref{LSE2}), where we approximate $E_{\bm k}\approx E_{\bm k}^{(0)}$with $E_{\bm k}^{(0)}$ given by Eq. (\ref{Ek0}).
Then, the first-order spectral function for the self-energy is expressible as
\begin{equation}
\varGamma_{\bm k}^{(1)}(\varepsilon)=-\pi\sideset{}{'}\sum_{\bm k_2\bm k_3}b_{\bm k\bm k_2\bm k_3}^2[1+f(E_{\bm k_2})+f(E_{\bm k_3})]\delta(\varepsilon+E_{\bm k_2}+E_{\bm k_3}).\label{vG1}
\end{equation}
The corresponding self-energy is given by Eq. (\ref{LSE}) as
\begin{equation}
\mathcal{S}_{\bm k}^{(1)}(i\varepsilon_n)=-\frac{1}{2}\sideset{}{'}\sum_{\bm k_2\bm k_3}b_{\bm k\bm k_2\bm k_3}^2\frac{1+f(E_{\bm k_2})+f(E_{\bm k_3})}{i\varepsilon_n+E_{\bm k_2}+E_{\bm k_3}}.\label{S1}
\end{equation}
Using Eq. (\ref{vG1}) and neglecting $\rm{Re}\mathcal{S}^{\rm{R}}(\varepsilon)$ in Eq. (\ref{A}), we obtain the first-order correction to the spectral function as
\begin{equation}\begin{aligned}
\mathcal{A}_{\bm k}^{(1)}(\varepsilon)\equiv&\frac{\varGamma_{\bm k}^{(1)}(\varepsilon)}{(\varepsilon-E_{\bm k})^2+[\varGamma_{\bm k}^{(1)}(\varepsilon)/2]^2}\\
\approx&-\pi\sideset{}{'}\sum_{\bm k_2\bm k_3}b_{\bm k\bm k_2\bm k_3}^2\frac{1+f(E_{\bm k_2})+f(E_{\bm k_3})}{(E_{\bm k}+E_{\bm k_2}+E_{\bm k_3})^2}\\
&\times\delta(\varepsilon+E_{\bm k_2}+E_{\bm k_3}),\label{A1}
\end{aligned}\end{equation}
where we have neglected $[\varGamma_{\bm k}^{(1)}(\varepsilon)/2]^2$ in the denominator.
Substituting this expression into Eq. (\ref{nkL}), we obtain
\begin{equation}\begin{aligned}
\braket{\hat{\gamma}_{\bm k}^\dag\hat{\gamma}_{\bm k}}=&\frac{1}{2}\sideset{}{'}\sum_{\bm k_2\bm k_3}\frac{b_{\bm k\bm k_2\bm k_3}^2}{(E_{\bm k}+E_{\bm k_2}+E_{\bm k_3})^2}\\
&\times[1+f(E_{\bm k_2})][1+f(E_{\bm k_3})].\label{nk1}
\end{aligned}\end{equation}
On the other hand, the zeroth-order expression of Eq. (\ref{A0}) may suffice to evaluate Eq. (\ref{n123L}) in the weak-coupling region, thus obtaining
\begin{equation}\begin{aligned}
\braket{\hat{\gamma}_{\bm k_1}\hat{\gamma}_{\bm k_2}\hat{\gamma}_{\bm k_3}}=&-\frac{b_{\bm k_1\bm k_2\bm k_3}}{E_{\bm k_1}+E_{\bm k_2}+E_{\bm k_3}}\\
&\times\left[\prod_{i=1}^3[1+f(E_{\bm k_i})]-\prod_{i=1}^3f(E_{\bm k_i})\right].\label{n1230}
\end{aligned}\end{equation}

We clarify how the results of the zero-temperature formalism are reproduced.\cite{3/2}
Taking the limit of $T\rightarrow0$ where $f(E_{\bm k})\rightarrow0$ and substituting Eq. (\ref{n1230}) with Eq.(\ref{b123}) into Eq. (\ref{chi}), we obtain the expression of $\chi_{\bm k}\equiv\lambda_{\bm k}u_{\bm k}^2$ that coincides with that of the zero-temperature formalism with the correspondence
\begin{equation}
w_{\bm k_1\bm k_2\bm k_3}=\braket{\hat{\gamma}_{\bm k_1}\hat{\gamma}_{\bm k_2}\hat{\gamma}_{\bm k_3}}=-\frac{b_{\bm k_1\bm k_2\bm k_3}}{E_{\bm k_1}+E_{\bm k_2}+E_{\bm k_3}}
\end{equation}
to the leading order. Eq. (\ref{nk1}) in the limit of $T\rightarrow0$ is identical to $\braket{\Phi|\hat{\tilde{\gamma}}_{\bm k}^\dag\hat{\tilde{\gamma}}_{\bm k}|\Phi}$ in the zero-temperature formalism.

If we introduce the Lehmann representation for Green's function in coordinate space, we can check the connection to the zero-temperature formalism in an inhomogeneous system\cite{3/2i} in the same manner.
That is, it is considered that our formalism is a natural extension of the zero-temperature formalism\cite{3/2,3/2m,3/2i}.
On the other hand, we note that Eq. (\ref{nk1}) cannot be used to solve the self-consistent equations because $\mathcal{A}_{\bm k}^{(1)}(\varepsilon)$ expressed by Eq. (\ref{A1}) does not satisfy the sum rule.

\bibliographystyle{jpsj}

\end{document}